\documentclass[a4paper,12pt]{article}
\usepackage{epsfig}
\usepackage{amsmath}
\usepackage{amssymb}

\begin{document}
\def\beq{\begin{equation}}
\def\eeq{\end{equation}}
\def\bea{\begin{eqnarray}}
\def\eea{\end{eqnarray}}
\def\ve{\vert}
\def\nnb{\nonumber}
\def\ga{\left(}
\def\dr{\right)}
\def\aga{\left\{}
\def\adr{\right\}}
\def\rar{\rightarrow}
\def\nnb{\nonumber}
\def\la{\langle}
\def\ra{\rangle}
\def\ba{\begin{array}}
\def\ea{\end{array}}

\title{Branching ratio of rare decay $B^0(B_s)\to \gamma\nu
\bar{\nu}$\thanks{This work is partly supported by National
Science Foundation of China.}}

\author{Jun-Xiao Chen$^{a,b}$, Zhao-Yu Hou$^c$, Ying Li$^a$ and Cai-Dian L\"u$^a$\\
 {\it \small $a$   Institute of High Energy Physics,
CAS, P.O.Box 918(4)  Beijing 100049, China}\\
{\it \small
$b$ Physics Department, Hebei Normal University, Shijiazhuang, Hebei 050016, China} \\
{\it \small$c$ Physics Department, Shijiazhuang Railway Institute,
Shijiazhuang, Hebei 050043, China} }

\maketitle
\begin{abstract}
The three-body decay $B^0(B_s)\to \gamma\nu \bar{\nu}$ can occur
via penguin and box diagrams in the standard model (SM). These
channels are useful to determine the decay constants $f_B$
($f_{B_s}$) and $B$ ($B_s$) meson wave function. Using the B meson
wave function determined in hadronic $B (B_s)$ decays, we
calculate and get the branching ratio of order $10^{-9}$ and
$10^{-8}$ for $B^0$ and $B_s$ decay, respectively. They agree with
previous calculations.
\end{abstract}

\section{Introduction}{\label{sec:intro}}

The flavor changing neutral current   process is one of the most
important field for testing the Standard Model (SM) at loop level
and for establishing new physics beyond that. The rare $B$ decays
provide a direct and reliable tool for extracting information
about the fundamental parameters of the Standard Model (SM), such
as, the Cabibbo-Kobayashi-Maskawa (CKM) matrix elements $V_{td}$
and $V_{ts}$, if we know the value of the decay constant $f_B$
from other methods. Conversely, we can determine the decay
constant $f_B$ if the CKM matrix elements are known.

Pure leptonic decays $B_s \to \mu^+ \mu^-$ and $B_s \to e^+e^-$,
are difficult to be measured in experiments, since helicity
suppression give a very small branching ratio at order
$\mathcal{O}(10^{-9})$ and $\mathcal{O}(10^{-14})$, respectively
\cite{8}. For $B^0$ meson case the situation gets even worse due
to the smaller CKM matrix elements $V_{td}$. For decay $B_s \to
\tau ^+\tau^-$, although its branching ratio is about $10^{-7}$
\cite{7}, it is still hard for experiments due to the low
efficiency of $\tau$ lepton measurements.

The $B^0(B_s ) \to \nu \bar \nu$ decay is forbidden due to
massless neutrino. Fortunately, having an extra real photon
emitted, the radiative leptonic decays can escape from the
helicity suppression, so that larger branching ratio of
$B^0(B_s)\to \gamma\nu \bar{\nu}$ is expected. A preliminary work
of this type decay was carried out with many different approaches
both in SM \cite{9604378,9610255,9710323} and beyond SM
\cite{0211159}. In above work, it was shown that the diagrams with
photon radiation from light quarks give the dominant contribution
to the decay amplitude, that is inversely proportional to the
constituent light quark mass. However the ``constituent quark
mass'' is poorly understood. In this work, we calculate the
branching ratio using $B$  meson wave function which describes the
constituent quark momentum distribution. The  wave function has
been studied for many years \cite{9902205} and used in calculating
non-leptonic of $B$ decay \cite{9411308}. Recently, this approach
is also used to calculate radiative leptonic decay of charged $B$
meson \cite{0505045}.

In the next section we analyze the relevant effective Hamiltonian
for the $B^0(B_s)\to \gamma\nu \bar{\nu}$ decay. In
Section.\ref{XXX}, we give our analytical and numerical results,
and then compare with other results. At last, we summarize this
work in section \ref{AAA}.

\section{Effective Hamiltonian}\label{sec:hami}

Let us first look at the quark level process $b  \to q \nu \bar
\nu$, with $q=s$ or $d$, which  is  shown in Fig.~\ref{fig1}. This
is a flavor changing neutral current   process, and both box and
$Z$ penguin contribute to this process. The effective Hamiltonian
in SM is given \cite{9340225}:

\begin{figure}[htbp]
\begin{center}
\includegraphics[scale=0.8]{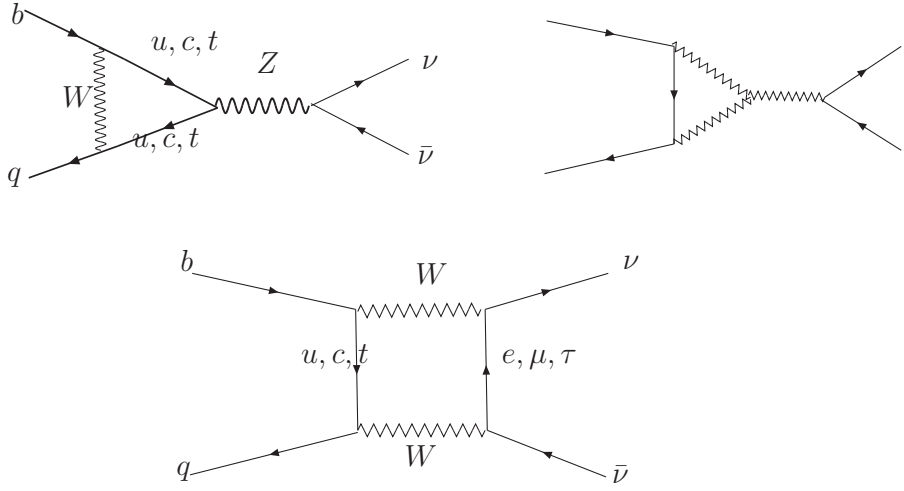}
\caption{Leading order Feynman diagrams in SM for $b \to q\nu
\bar\nu$, with $q=s$ or $d$.} \label{fig1}
\end{center}
\end{figure}

\begin{equation}\label{Hamilton}
H=C(\bar{q}\gamma_{\mu}P_Lb)(\bar{\nu}\gamma^\mu P_L\nu),
\end{equation}
with $P_L=(1-\gamma_5)/{2}$. The coefficient $C$ is
\begin{equation}\label{Coefficint}
C=\frac{\sqrt{2}G_F\alpha}{\pi\sin^2\theta_w}V_{tb}V^*_{tq}
\frac{x}{8}\left[\frac{x+2}{x-1}+\frac{3x-6}{(x-1)^2}\ln{x}\right],
\end{equation}
and $x={m_t^2}/{m_W^2}$. From this expression, we can see that the
coefficient $C$ is sensitive to the mass of the particle in loop.
If new particle exist, it should affect the Wilson Coefficient and
change the branching ratio. That is why this kind of  flavor
changing neutral current processes is sensitive to new physics
\cite{0211159}.

We have already mentioned that the pure leptonic ($\nu\bar\nu$)
decay is forbidden due to helicity conservation. However, when a
photon is emitted from any charged line of $b$ or $q$ quark, this
pure leptonic processes turn into radiative ones and   helicity
suppression does not exist anymore. At quark level the process
$B_{s(d)} \rar \gamma\nu \bar \nu $ is described by the same
diagrams as $b \rar q \gamma \nu \bar \nu$ shown in
Fig.\ref{fig2}. Incidentally, we should note the following
peculiarities of this process:

\begin{figure}[htbp]
\begin{center}
\includegraphics [scale=0.8] {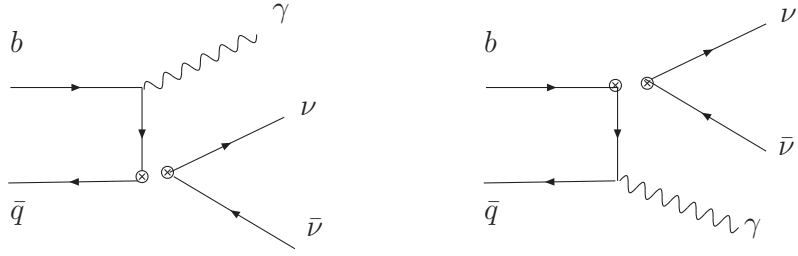}
\caption{Feynman diagrams  for $b \to q\gamma\nu  \bar\nu$ using
effective four fermi operators.} \label{fig2}
\end{center}
\end{figure}

\begin{itemize}
    \item when photon emitted from
          internal charged particles (W or top quark), the above mentioned
          process will be suppressed by a factor $ {m_b^2}/{m_W^2}$ (see
          \cite{9604378}), in comparison to the process $b \to q \nu \bar \nu$,
          one can neglect the contribution of such diagrams.
    \item The Wilson coefficient $C$ is the same for the processes $b
          \to q \gamma\nu \bar \nu $ and $b \rar q \nu \bar \nu$ as a
         consequence of the extension of the Low's low energy theorem (for
         more detail see \cite{902314}).

\end{itemize}

So, when photon emitted from initial $b$ or light quark line,
there are only two   diagrams,   contributing to the process $b
\rar q \gamma\nu \bar \nu $. From Fig.\ref{fig2}, the
corresponding decay amplitude turns out to be
\begin{equation}
{\cal A} = -\frac{e}{6} C ~\bar q \left [ \not \!
\epsilon^*_\gamma \frac{ \not \!p _\gamma -\not \! p_q +m_q}{(p_q
\cdot p_\gamma)} \gamma_\mu P_L+P_R \gamma_\mu \frac{\not \! p_b
-\not \! p_\gamma +m_b}{(p_b\cdot p_\gamma)} \not \!
\epsilon^*_\gamma \right] b ~(\bar \nu \gamma ^\mu P_L \nu)
\label{h4}.
\end{equation}

\section{Analytical and Numerical
results}\label{XXX}

In order to calculate analytic formulas of the decay amplitude, we
use the wave functions $\Phi_{M, \alpha\beta}$ decomposed in terms
of spin structure. In the summation procedures, the $B$ meson is
treated as a heavy-light system. Thus, the $B$ meson light-cone
matrix element can be decomposed as \cite{013592}:
\begin{equation}
 \Phi_{B,\alpha\beta} = \frac{i}{\sqrt{2N_c}}
\left\{ (\not \! P_B \gamma_5)_{\alpha\beta} \phi_B^A +
\gamma_{5\alpha\beta} \phi_B^P \right\},
\end{equation}
where $N_c = 3$ is color degree of freedom, $P_B$ is the
corresponding meson's momentum,  $\phi_B^{A}$ and $\phi_B^{P}$ are
Lorentz scalar distribution amplitudes. As heavy quark effective
theory leads to $\phi_B^P \simeq M_B \phi_B^A$, then $B$ meson's
wave function can be expressed by
\begin{equation}
 \Phi_{B,\alpha\beta}(x) = \frac{i}{\sqrt{2N_c}}
\left[ \not \! P_B  + M_B \right]\gamma_{5\alpha\beta} \phi_B(x).
\end{equation}
In above function, the function $\phi_B$ describes the momentum
distribution amplitude. Since $b$ quark is much heavier than the
light quark in  $B$ meson,   there is a sharp peak at the small
$x$ region for the light quark momentum fraction,
\begin{equation}
\phi_B(x) = N_B x^2(1-x)^2 \exp \left[ -\frac{M_B^2\ x^2}{2
\omega_b^2}  \right]. \label{phib}
\end{equation}
It satisfies the normalization relation:
\begin{equation}
\int_0^1\phi_B(x)dx=\frac{f_B}{2\sqrt{2N_c}}\;, \label{no}
\end{equation}
with $f_B$ the $B$ meson decay constant. This choice of $B$
meson's wave function is almost a best  fit from the $B$ meson
non-leptonic two body decays \cite{9411308}.

For simplicity, we consider the B meson at rest and use the
 light-cone coordinate $(p^+,p^-,\vec p_\perp)$   to describe the
 meson  and quark's momenta, where $p^\pm=\frac{1}{\sqrt{2}}(p^0\pm
 p^3)$ and $p_\perp=(p^1,p^2)$. Using these coordinate we can take
 the $B_q$, $\nu \bar {\nu}$ (momentum sum) and photon's momenta as
 \begin{equation}
 P_B=\frac{M_B}{\sqrt{2}}(1,1,\vec{0}_\perp);~\ \ ~ P_{\nu
 \bar{\nu}}=\frac{M_B}{\sqrt{2}}(1,r^2,\vec{0}_\perp); ~\ \ ~
P_\gamma=\frac{M_B}{\sqrt{2}}(0,1-r^2,\vec{0}_\perp),
 \end{equation}
with $r^2= {P_{\nu \bar{\nu}}^2}/{M_B^2}$. The momenta of $b$ and
$q$ quark in B meson are  $p_b=(1-x) P_B$, $p_q=xP_B$. Using above
convention, the amplitude for $B_q\to \gamma \nu \bar {\nu}$ decay
is written by:
\begin{equation}
\label{amplitude} A=\frac{\sqrt{6}eC}{6}
\left[~iC_1\epsilon_{\alpha\beta\mu\nu}\varepsilon^{*\alpha}
P_\gamma^\beta P_B^\nu
+C_2~(P_{\gamma\mu}\varepsilon^*_\nu-P_{\gamma\nu}\varepsilon^*_\mu)~P_B^\nu\right]
(\bar{\nu_1} \gamma^\mu p_L \nu_2)
\end{equation}
with
\begin{eqnarray}\label{constants1}
C_1=\int_0^{1} \frac{\Phi(x)}{p_b \cdot
P_\gamma}+\int_0^{1}\frac{\Phi(x)}{p_q \cdot P_\gamma} \\
C_2=\int_0^{1}\frac{\Phi(x)}{p_q \cdot,
P_\gamma}-\int_0^{1}\frac{\phi(x)}{p_b \cdot P_\gamma}.
\end{eqnarray}

 After squaring the amplitude and performing the phase
space integration over one of the two Dalitz variables, and
summing over three generation of neutrinos, we get the
differential decay width versus the photon energy $E_\gamma$:
\begin{eqnarray}\label{distribution}
\frac{d\Gamma}{dE_\gamma}=\frac{6 C^2\alpha}{(12\pi)
^2}~(C_1'^2+C_2'^2)~(M_B-2E_\gamma)E_\gamma ,
\end{eqnarray}
with
\begin{eqnarray}
C_1'^2=\left( \int_0^1
\frac{\Phi(x)}{1-x}dx+\int_0^1\frac{\Phi(x)}{x}dx\right)^2 \\
C_2'^2=\left( \int_0^1
\frac{\Phi(x)}{1-x}dx-\int_0^1\frac{\Phi(x)}{x}dx\right)^2.
\end{eqnarray}
By integrating the variable $E_\gamma$, we get the decay width:
\begin{eqnarray}\label{decay}
\Gamma=\frac{M_B^3 C^2\alpha}{(24\pi)^2}(C_1'^2+C_2'^2).
\end{eqnarray}

 In this work, we use following
parameters \cite{9902205,pdg}:
\begin{eqnarray}
\omega_b=0.4,~ f_{B}=0.19~ \mathrm{GeV},
~\tau_{B^0}=1.54\times10^{-12}s;\nonumber\\
\omega_{b_s}=0.5, ~f_{Bs}=0.24~ \mathrm{GeV},
~\tau_{B_s}=1.46\times10^{-12}s;\nonumber\\
G_F=1.66\times 10^{-5}\mathrm{GeV}^{-2}; ~
\sin^2\theta_\omega=0.23; ~\alpha=\frac{1}{132} \nonumber \\
V_{tb}=0.999, ~~V_{td}=0.0074, ~~V_{ts}=0.041.
\end{eqnarray}
Using these parameters, we get the branching ratios:
\begin{eqnarray}
\mathbf{Br}(B^0\to \gamma\nu \bar{\nu})&=&0.7\times10^{-9},\nonumber\\
\mathbf{Br}(B_s\to \gamma\nu \bar{\nu})&=&2.4\times10^{-8}.
\end{eqnarray}

 Just as we mentioned above, many
approaches have been used to analyze these processes such as
constituent quark model \cite{9604378}, pole model
\cite{9604378,991199}, QCD sum rule \cite{9610255}, and light
front approach \cite{9710323}. Here we compare our results with
them which is shown in Table.~\ref{table}.

\begin{table}[htbp]
\caption{Comparison of results from different approaches}
\begin{center}
\begin{tabular}{|c|c|c|c|c|c|}
\hline\hline Mode & Our Results  & Quark Model & Pole Model& Sum
Rule &light
front\\
\hline
 $\mathbf{BR}(B^0\to \gamma\nu \bar{\nu})$
              & $ 0.74 \times 10^{-9}$
              & $ 1.7 \times 10^{-9}$
              & $ 2.1 \times 10^{-9}$
              & $ 4.2 \times 10^{-9} $
             & $ 1.4 \times 10^{-9} $
              \\
\hline
 $\mathbf{BR}(B_s\to \gamma\nu \bar{\nu})$
              & $2.4\times 10^{-8}$
              & $3.5\times 10^{-8}$
              & $1.8\times 10^{-8}$
              & $7.5\times 10^{-8}$
              & $2.0\times 10^{-8}$
               \\
 \hline
\end{tabular}\label{table}
\end{center}
\end{table}

In constituent quark model, the non-relativity character is
considered. If we replace our $B$ meson distribution amplitude in
eq.(\ref{phib}) by a $\delta$ function $({f_B}/{2\sqrt{6}}) ~
\delta(x-m_q/m_B)$, our formula will   return back to the
constituent quark model in ref.\cite{9604378}.   Since we have
poor knowledge about quark mass up to now, our new calculation is
surely an improvement. In paper \cite{9604378}, the authors also
calculate these processes in the pole model, and the results are
similar to the quark model case. From the table, one can also see
that most of the methods get similar results except the QCD sum
rule approach, whose result  is  larger than others.   We hope the
experiments in future can test these different methods.

\begin{figure}[htbp]
\begin{center}
\includegraphics[scale=1]{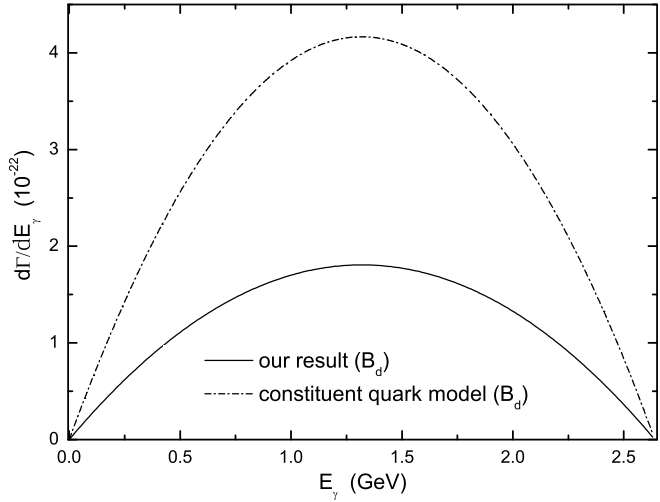}
\caption{Differential decay rate of $B^0 \to \gamma \nu\bar\nu$
versus the photon energy $E_{\gamma}$.}\label{fig3}
\end{center}
\end{figure}

\begin{figure}[htbp]
\begin{center}
\includegraphics[scale=1]{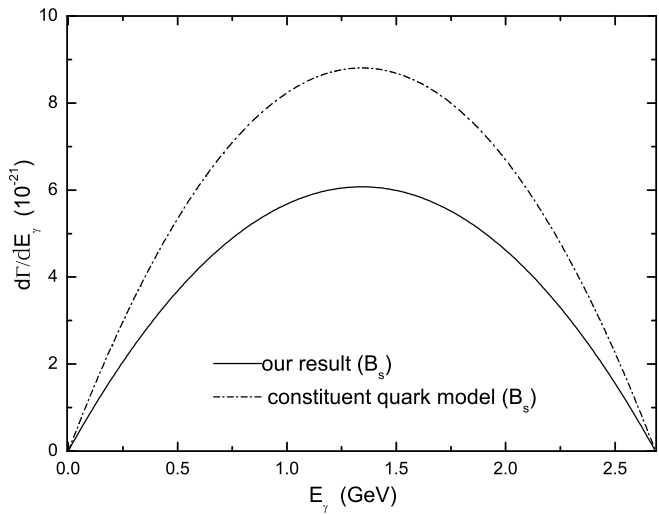}
\caption{Differential decay rate of $B_s \to \gamma \nu\bar\nu$
versus the photon energy $E_{\gamma}$.} \label{fig4}
\end{center}
\end{figure}

 In Fig.~(\ref{fig3}) and
Fig.~(\ref{fig4}), we figure out the differential decay rate of
$B^0 (B_S) \to \gamma \nu\bar\nu$ versus photon energy $E_\gamma$.
We also display the photon energy spectrum  from constituent quark
model\footnote{The results of constituent quark model  are from
updated parameters.}. From these figures, we find our results are
smaller than the constituent quark one, but the shape of the
spectrum is the same. If normalized decay rate is used, the two
lines will become only one, since the function is very simple
\begin{equation}
f(x) = 24 x(1-2x)
\end{equation}
 which can be extracted from
eq.(\ref{distribution}).

Of course, there are  also uncertainties in our calculation. The
most large uncertainty comes from the heavy meson wave function.
The high order twists contribution, the high Fock states are also
not included, because they are not clear completely now. We hope
that the non-leptonic $B$ meson decay can offer more information
  in the near future.

\section{Summary}\label{AAA}

In this work, we calculate the branching ratios in SM for $B_s\to
\gamma\nu \bar{\nu}$ to be $10^{-8}$ and for $B^0\to \gamma\nu
\bar{\nu}$ to be $10^{-9}$ using $B$ meson wave function
constrained from non-leptonic B decays. These decay channels are
useful to determine the decay constants $f_B$ and wave function.
After calculation, we find our leading order results are at the
same order as other approaches but a little smaller. These rare
decays are sensitive to any new physics contributions which can be
measured by future experiment such as LHCb.

\section*{Acknowledgements} We thank Y-L. Shen and J. Zhu for
the help on the program, and also thank X-Q. Yu and W. Wang for
helpful discussions.

\end{document}